\documentclass[twocolumn]{aastex631}
\usepackage[T1]{fontenc}
\usepackage{amsmath}
\usepackage{amssymb}
\usepackage{soul}
\usepackage{graphicx} 
\usepackage{subfigure}
\begin{document}

\title{Choked precessing jets in tidal disruption events and high-energy neutrinos}

\author[0009-0003-7748-3733]{Qi-Rui Yang}
\affiliation{School of Astronomy and Space Science, Nanjing University, Nanjing 210023, People’s Republic of China}
\affiliation{Key laboratory of Modern Astronomy and Astrophysics (Nanjing University), \\
Ministry of Education, Nanjing 210023, People’s Republic of China}

\author[0000-0001-5751-633X]{Jian-He Zheng}
\affiliation{School of Astronomy and Space Science, Nanjing University, Nanjing 210023, People’s Republic of China}
\affiliation{Key laboratory of Modern Astronomy and Astrophysics (Nanjing University), \\
Ministry of Education, Nanjing 210023, People’s Republic of China}

\author[0000-0003-1576-0961]{Ruo-Yu Liu}
\affiliation{School of Astronomy and Space Science, Nanjing University, Nanjing 210023, People’s Republic of China}
\affiliation{Key laboratory of Modern Astronomy and Astrophysics (Nanjing University), \\
Ministry of Education, Nanjing 210023, People’s Republic of China}

\author[0000-0002-5881-335X]{Xiang-Yu Wang}
\affiliation{School of Astronomy and Space Science, Nanjing University, Nanjing 210023, People’s Republic of China}
\affiliation{Key laboratory of Modern Astronomy and Astrophysics (Nanjing University), \\
Ministry of Education, Nanjing 210023, People’s Republic of China}
\email{xywang@nju.edu.cn}

\begin{abstract}
It has been suggested that relativistic jets might have
been commonly formed in tidal disruption events (TDEs), but those with relatively weak power could be choked by the surrounding envelope. The discovery of high-energy neutrinos possibly associated with some normal TDEs may support this picture in the hypothesis that the neutrinos are produced by choked jets. Recently, it was noted that disrupted stars generally have misaligned orbits with respect to the supermassive black hole  spin axis and highly  misaligned precessing jets are more likely to be choked. Here we revisit the jet break-out condition for misaligned precessing jets by considering the jet could be collimated by the cocoon pressure while propagating in the disk wind envelope. The jet head opening angle decreases as the jet propagates in the envelope, but the minimum power of a successful jet remains unchanged in terms of the physical jet power. We further calculate the neutrino flux from choked precessing jets, assuming that the cocoon energy does not exceed the kinetic energy of the disk wind. We find that neutrino flux from highly  misaligned choked jets is sufficient to explain the neutrinos from AT2019aalc, while it is marginal to explain the  neutrinos from AT2019dsg and AT2019fdr. 
The latter could be produced by weakly  misaligned choked jets, since the  duty cycle  that the
jet sweeps across increases as the misaligned angle decreases. 
We also show that  the population of choked TDE jets  could contribute to $\sim 10\%$ of the observed diffuse neutrino flux measured by IceCube.

\end{abstract}

\section{Introduction}
A couple of TDEs with relativistic jets (hereafter “jetted TDEs”) have
been observed, including Swift J1644+57 \citep{2011Sci...333..203B}, Swift J2058+05 \citep{2012ApJ...753...77C}, Swift J1112+82 \citep{2015MNRAS.452.4297B}, and AT2022cmc
\citep{2022Natur.612..430A}. The volumetric rate of  on-axis  jetted TDEs  has been inferred
to be ${\rm \sim 0.03 Gpc^{-3} yr^{-1}}$   \citep{Burrows2011Natur.476..421B,2022Natur.612..430A}. If we correct for a beaming
factor $f_b\sim 0.01$ (adopting a typical  half-opening angle of $\theta_{\rm j}\sim 0.1$ ), the inferred intrinsic rate of jetted TDEs is   ${\rm \sim 3 Gpc^{-3} yr^{-1}}$. This is a very small
fraction ($\sim 1\%$) of the rate of all TDEs observed in various
surveys, $R\sim 300 {\rm Gpc^{-3} yr^{-1}}$  \citep{2018ApJ...852...72V,2021MNRAS.508.3820S,Yao2023ApJ...955L...6Y}. The isotropic equivalent jet  luminosity of these jetted TDEs is in the range of $\sim 10^{47}-  10^{48}  ~{\rm erg s^{-1}}$.  Since the black hole
accretion is expected to produce a continuous distribution of jet
luminosity, as has been seen in  active galactic nuclei (AGN) jets, the fact that no less
powerful  jets have been found in  TDEs  so far is a puzzle.

\cite{Wang2016PhRvD..93h3005W}  suggested that, while powerful jets, such as that in Swift J1644+57, can successfully break
out from the surrounding envelope, weak jets could be choked during the propagation inside the envelope. This could explain why no less
powerful  jets have been found in  TDEs.  Like  successful jets,  the choked jet can also accelerate cosmic rays through  shocks and further produce neutrinos via
interaction with the surrounding dense photons \citep{Wang2011PhRvD..84h1301W,Wang2016PhRvD..93h3005W}. Recent claims that
high-energy neutrino events detected by IceCube are coincident
with three TDE candidates (AT2019dsg, AT2019fdr and
AT2019aalc)  stimulated the study on neutrino production in TDEs.   Various production sites of high-energy neutrinos
have been proposed, including successful jets\citep{Liu2020PhRvD.102h3028L,Winter2021NatAs...5..472W}, choked  jets \citep{Zheng2023ApJ...954...17Z,Mukhopadhyay2023arXiv230902275M}, disk outflow \citep{2020ApJ...902..108M,wu2022MNRAS.514.4406W, 2023ApJ...948...42W}, and the AGN core \citep{2019ApJ...886..114H,2020ApJ...902..108M}. In particular, \cite{Zheng2023ApJ...954...17Z} explore the parameter space of the jets that can produce detectable neutrino flux while being choked in the
expanding envelope, finding that the cumulative neutrino numbers of AT2019fdr and AT2019aalc are consistent
with the expected range imposed by observations, although the allowed parameter space for AT2019dsg is small. \cite{Mukhopadhyay2023arXiv230902275M} discussed high-energy neutrino production in
delayed choked jets, and find that the  a delay of the jet launch  can make  the jets more easily choked and  can increasing the allowed neutrino flux. 

Very recently, \cite{Teboul2023ApJ...957L...9T} and \cite{Lu2023arXiv231015336L} noted that disrupted stars generally have misaligned orbits with respect to the supermassive black hole (SMBH) spin axis and  the accretion
disk fed by the fallback stellar debris undergoes Lense-Thirring precession around the BH spin axis. They found that that precessing jets can only break out of the wind envelope confinement if
the misalignment angles $\theta_{\rm LS}$ (between the jet axis and the BH spin axis) is less than a few times the jet opening angle $\theta_{\rm j}$, explaining the tiny fraction of  jetted TDEs.  In most TDEs with $\theta_{\rm LS}\gg \theta_{\rm j}$, the jets are initially choked by the disk wind and may only break out later when
the disk finally aligns itself with the spin axis due to the  damping of the precession \citep{Teboul2023ApJ...957L...9T,Lu2023arXiv231015336L}.

In this paper, we study the neutrino production in choked TDE jets when the jet is precessing. \cite{Teboul2023ApJ...957L...9T} and \cite{Lu2023arXiv231015336L} both analytically  study the propagation of a
misaligned precessing jet inside the disk wind.  \cite{Lu2023arXiv231015336L} also carry out three-dimensional relativistic hydrodynamic
simulations of a precessing jet. In \S 2, we first revisit  the jet propagation and the break-out condition for misaligned precessing jets by considering that the jet could be collimated by the cocoon pressure while it is propagating in the envelope. In \S 3, we calculate the neutrino flux produced by choked precessing jets and compare with the observations of three TDEs. We also calcuate the diffuse neutrino emission from the population of choked TDEs.  Finally we give  discussions and conclusions in \S 4.

\section{Jet propagation and the break-out condition for   precessing jets }
After the stellar disruption,  an accretion disk is formed in the orbital
plane of the star whose normal is generally  misaligned by an angle $\theta_{\rm LS}$ with respect to the SMBH spin axis. The
accretion disk undergoes Lense-Thirring precession around the black hole spin axis and  the jet is aligned with the instantaneous disk angular momentum \citep{Teboul2023ApJ...957L...9T,Lu2023arXiv231015336L}. Matter falls back to the disk at a highly super-Eddington accretion rate $\dot{M}$, driving a disk wind of velocity $\beta_{\rm w}=v_{\rm w}/c$ and mass-loss rate $\dot{M}_{\rm w}=f_{\rm w} \dot{M}$, where $f_{\rm w}<1$ is the fraction of the mass fallback rate lost in the wind.     \cite{Metzger2016MNRAS.461..948M} argued that only a small fraction the returning debris ends up reaching the inner circular disk and accreting onto the SMBH, with  the vast majority  becoming unbound in a wind outflow. So  we consider $f_{\rm w} = 0.5$ in our calculation.  We choose $\beta_{\rm w} = 0.1$ as a typical wind velocity.
Due to the rapid precession, these disk winds will inevitably interact and
collide with each other on large scales, leading to   a quasi-spherical outflow.

{We consider a relativistic jet with a kinetic luminosity $L_{\rm j}$,  an initial half-opening
angle $\theta_0$ and
 a  bulk Lorentz factor  $\Gamma=10\ga 1/\theta_0$}.
As the jet advances in the quasi-spherical wind envelope, the jet
drives a bow shock ahead of it. The jet is capped by a termination shock, and a reverse shock propagates back into the jet, where the jet is decelerated and heated. The hot head material spills sideways, forming a cocoon that engulfs the jet and collimate it.
The velocity of the jet head is determined by the balance of pressure between ambient gas and the jet head \citep{Matzner2003MNRAS.345..575M}. In a moving wind envelope, we can express the velocity in the frame of
the wind envelope by a coordinate transformation  \citep[e.g.,][]{Ioka2018PTEP.2018d3E02I,Hamidani2020MNRAS.491.3192H,Gottlieb2022MNRAS.517.1640G,Lu2023arXiv231015336L}:
\begin{equation}
\beta_{\rm h}-\beta_{\rm w}\simeq \frac{\beta_{\rm j}-\beta_w}{1+\tilde{L}^{-1/2}}, 
\label{betahead}
\end{equation}
where 
\begin{equation}
  \tilde{L}=  \frac{L_{\rm j}}{\Sigma_{\rm j} \rho_{\rm w}(r_{\rm h})c^3} = \frac{2L_{\rm j,iso} \beta_{\rm w} }{\dot{M}_{\rm w} c^2}.
  \label{Ltilde}
\end{equation}
Here $\beta_{\rm h}$ is the jet head velocity in unit of the speed of light (measured in the lab frame), $r_{\rm h}$ is the radius of the jet head,  $\rho_{\rm w}$ is wind density  and $\Sigma_{\rm j}=\pi r^{2}_{\rm h} \theta^{2}_{\rm j}$ is the jet cross-section at the location of the head with $\theta_{\rm j}$ being the jet head opening angle\footnote{The jet head opening angle is defined as $\theta_{\rm j} = r_{\rm j}/r_{\rm h}$ (regardless of the jet collimation), where $r_{\rm j}$ is the radius of jet head cross-section and $r_{\rm h} = \beta_{\rm h}ct$ is the height of the jet head.}. $L_{\rm j,iso}=2L_{\rm j} / \theta_{\rm j}^2$ is the isotropic equivalent luminosity and $\dot{M}_{\rm w} = 4\pi r^{2}_{\rm h} \rho_{\rm w} \beta_{\rm w} c$  is disk wind mass-loss rate. 

\begin{figure}
\subfigure[]{
    \centering
    \includegraphics[width=0.5\textwidth]{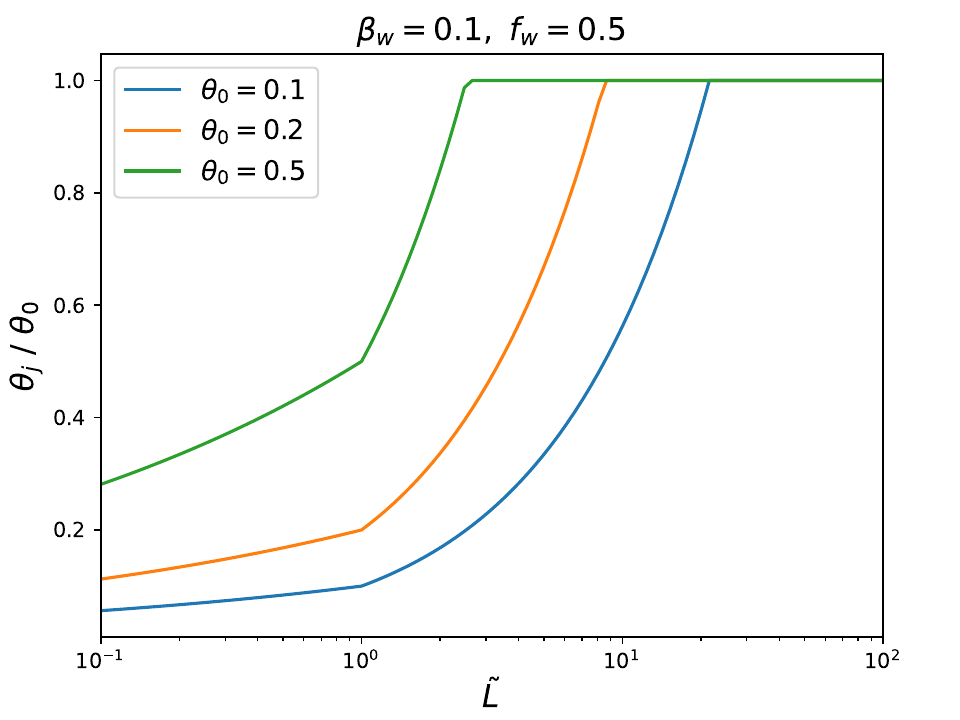}
    }
\subfigure[]{
    \centering
    \includegraphics[width=0.5\textwidth]{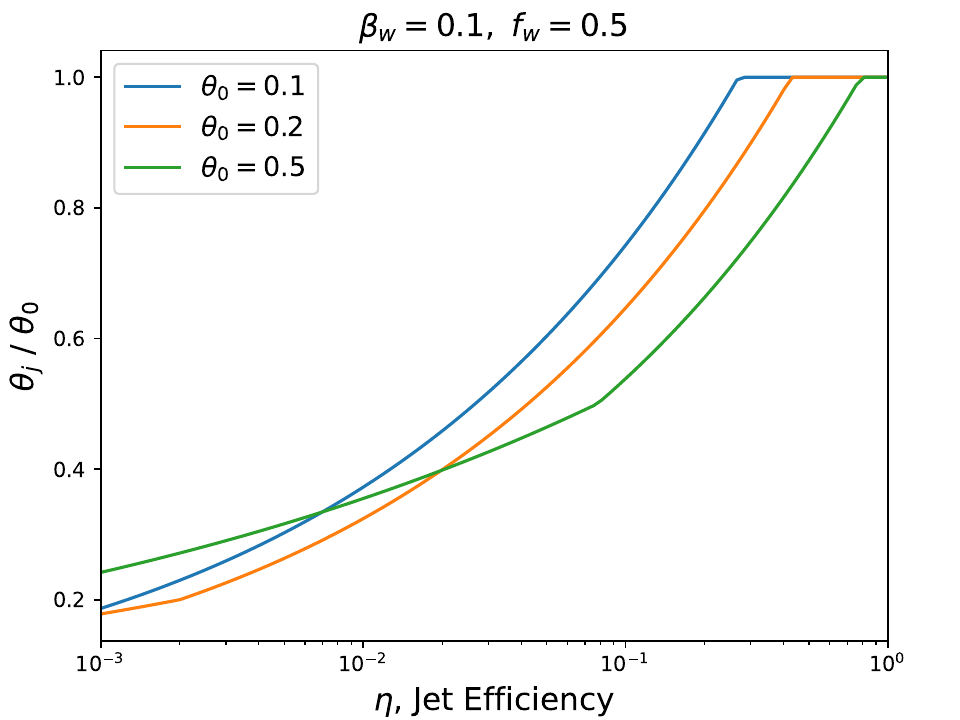}
    }
\caption{The evolution of $\theta_{\rm j}$ adopting Eq.\eqref{Thetaj}. Top: $\theta_{\rm j} / \theta_0$ as a function of $\tilde{L}$ for various initial jet opening angles $\theta_0$. Bottom: $\theta_{\rm j} / \theta_0$ as a function of the jet efficiency $\eta$ for various initial jet opening angles $\theta_0$.}
    \label{thetaj evo} 
\end{figure}

Due to precession, the jet becomes episodic   in a given direction. For such an episodic
jet, if the timescale  for the jet head to break out from the wind envelope is longer than the active timescale of the jet episode, the jet is choked \citep{Lu2023arXiv231015336L}. Following \cite{Lu2023arXiv231015336L}, we consider that, along a fixed direction, the jet is “on” for a duration
of $t_{\rm on}$ and “off” for a duration of $T_{\rm prec}-t_{\rm on}$, where $T_{\rm prec}$ is the
precessional period. The jet head will catch up with the outer edge of the wind
after a  time
\begin{equation}
t_{\rm bo}=\frac{\beta_{\rm w} (T_{\rm prec}-t_{\rm on})}{\beta_{\rm h} - \beta_{\rm w}}.
\label{tbo}
\end{equation} 
The active timescale of the jet episode is the time for the  reverse shock 
crossing the jet episode, which is \citep{Lu2023arXiv231015336L}
\begin{equation}
t_{\rm act}= \frac{\beta_{\rm j} t_{\rm on}}{\beta_{\rm j}- \beta_{\rm h}}.
\label{tcross}
\end{equation}

A successful jet break-out requires  $t_{\rm bo}/t_{\rm act} < 1$,  which
can be converted to the following form 
\begin{equation}
    \frac{L_{\rm j,iso}}{\beta_{\rm w} \dot{M}_{\rm w} c^2} > (\frac{1-\xi_{\rm duty}}{\xi_{\rm duty}})^{2},
    \label{boc1}
\end{equation}
where $\xi_{\rm duty} = t_{\rm on}/T_{\rm prec}$ is the duty cycle representing the fraction that the jet is "on" along a fixed direction of the whole precessing period, which can be expressed as $\xi_{\rm duty} \approx {\theta_{\rm j}/(\pi \rm{sin}\theta_{\rm LS}})$  when $\theta_{\rm LS}\gg \theta_j$ \citep{Lu2023arXiv231015336L}. Thus the break-out condition   is \citep{Lu2023arXiv231015336L}
\begin{equation}
L_{\rm j,iso}\gtrsim\beta_{\rm w} \dot{M}_{\rm w} c^2 \frac{\pi^2 \rm sin^2 \theta_{\rm LS}}{\theta_{\rm j}^2}.
\label{boc2}
\end{equation}

Jets may be collimated by the cocoon pressure  while it is propagating in the wind envelope. For the precession jet in our case, we assume that during the "on" period, the jet propagation is almost straight and the wind envelope is quasi-steady. The critical parameter $\tilde{L}$ determines evolution of the jet \citep{Bromberg2011ApJ...740..100B}. The jet is  collimated if the collimation shock converges below the jet head, which occurs when $\tilde{L} < \theta_0^{-4/3}$.   The  opening angle of the jet head is thus given by \citep{Bromberg2011ApJ...740..100B}
\begin{align}
\begin{split}
\theta_{\rm j}= \left \{
\begin{array}{ll}
    \tilde{L}^{1/4}\theta_0^2 = (\frac{4 \eta \beta_{\rm w}}{f_{\rm w}})^{1/6} \theta_0^{4/3},                    & \tilde{L} < 1\\
    \tilde{L}^{3/4}\theta_0^2 = (\frac{4 \eta \beta_{\rm w}}{f_{\rm w}})^{3/10}\theta_0^{4/5},                    & 1< \tilde{L} < \theta_0^{-4/3}\\
    \theta_0       & \tilde{L}>\theta_0^{-4/3}
\end{array}
\right.
\end{split}
\label{Thetaj}
\end{align}
where  $\eta=L_{\rm j}/\dot{M}c^2$ is the jet efficiency.    Fig.\ref{thetaj evo} shows the evolution of $\theta_{\rm j}$ as a function of the dimensionless jet luminosity $\tilde{L}$ and the jet efficiency $\eta$. The upper panel shows that the dimensionless parameter $\tilde{L}$ can determine the jet head opening angle and velocity of jet head. For $\tilde{L} < 1$, the jet head is non-relativistic and vice versa. The lower panel indicates that for a low jet efficiency, the jet head opening angle is significantly lower than the initial opening angle of the jet as the result of the collimation. 

We would like to stress that due to the collimation, the jet head opening angle $\theta_{\rm j}$ is smaller than the initial jet opening angle $\theta_0$, and as a result, $L_{\rm j,iso}$ may change due to collimation. So we express the break-out condition using the beam-corrected jet luminosity
\begin{equation}
L_{\rm j}\ge \frac{1}{2}\dot{M} f_{\rm w} c^2 \beta_{\rm w}\pi^2 {\rm sin}^2\theta_{\rm LS}.
\label{bocj}
\end{equation}
This condition is independent of the jet  opening angle and holds regardless of whether the collimation effect is taken into account. This is because a decreasing opening angle $\theta_{\rm j}$ leads to a higher  $L_{\rm j,iso}$ and also a shorter jet activity period $t_{\rm act}$ due to jet precession, which cancel with each other.

\begin{figure}
    \centering
    \includegraphics[width=0.5\textwidth]{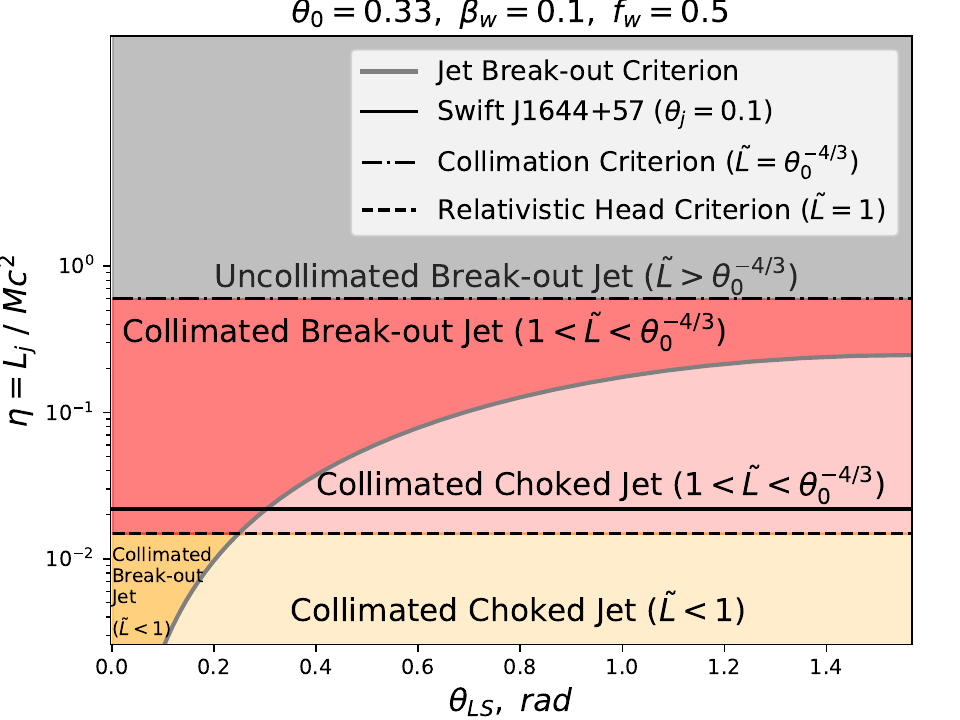}
    \caption{Break-out condition in terms of the jet efficiency $  L_{\rm j} / \dot{M} c^2$ and misaligned angle $\theta_{\rm LS}$ for the parameter values of $\rm \theta_0 = 0.33$, $ f_{\rm w} = 0.5$, $\beta_{\rm w} = 0.1$. The gray solid line shows the jet break-out criterion. The darker regions indicate break-out jet, whereas the lighter regions indicate choked jet. The yellow regions, the red regions and the gray region represent collimated jets with non-relativistic jet head, collimated jets with relativistic jet head and uncollimated jets, respectively. The horizontal black solid line shows the case of Swift J1644+57 with a jet efficiency of  $\eta = 0.023$.     Note that we simply just use the region larger than unity to denote the figure labels.} 
    \label{Lj} 
\end{figure}

The wind speed $v_{\rm w}$ is roughly the  local Keplerian speed of the accretion disk at the launching radius of the disk wind. For typical TDEs of a solar-like star  by a $10^6-10^7 M_\odot$ SMBH, $v_{\rm w}\simeq 0.1 {\rm c}$ \citep{Metzger2016MNRAS.461..948M,Lu2023arXiv231015336L}. The discovery of highly blueshifted ($v=0.05$c) broad Balmer and metastable helium absorption lines 
in   the optical spectra of  the TDE AT2018zr is consistent with this wind speed \citep{Hung2019ApJ...879..119H}. 
We show the break-out condition in Fig.\ref{Lj} for an example jet with $\theta_0=1/3$,  $f_{\rm w}=0.5$ and $\beta_{\rm w}=0.1$.   From top to the bottom (corresponding to a decreasing $L_{\rm j}/\dot{M} c^2$), the various color regions in Fig.\ref{Lj} represent the  uncollimated break-out jet (grey region), collimated break-out jet with a a relativistic jet head (red region),  the collimated choked jet with a relativistic jet head (pink region),
the collimated break jet with a sub-relativistic jet head (dark yellow region), the collimated choked jet with a sub-relativistic jet head (yellow region). Eq.\eqref{bocj} implies that break-out condition is independent of $\theta_0$ or $\theta_{\rm j}$. With fixed $f_{\rm w}$, $\beta_{\rm w}$ and misaligned angle $\theta_{\rm LS}$, the jet efficiency $\eta=L_{\rm j}/\dot{M}c^2$ entirely determines whether the precessing jet breaks out or not.

We show the position of a jetted TDE like Swift J1644+57 in Fig.\ref{Lj}.
The luminosity of  Swift J1644+57 is about $10^{48}~{\rm erg~s^{-1}}$ and the opening angle of the jet is assumed to be $\theta_{\rm j} = 0.1$. For a wind mass-loss rate  $2M_{\odot}{\rm yr^{-1}}$ \citep{Lu2023arXiv231015336L}, we derive $\eta = 0.023$. 
Swift J1644 can only have a successful jet break-out provided that the misalignment angle is sufficiently small ($\theta_{\rm LS} \lesssim 15^{\circ}$), consistent with the result in \citet{Lu2023arXiv231015336L}. 
 
\begin{figure}
    \centering
    \includegraphics[width=0.5\textwidth]{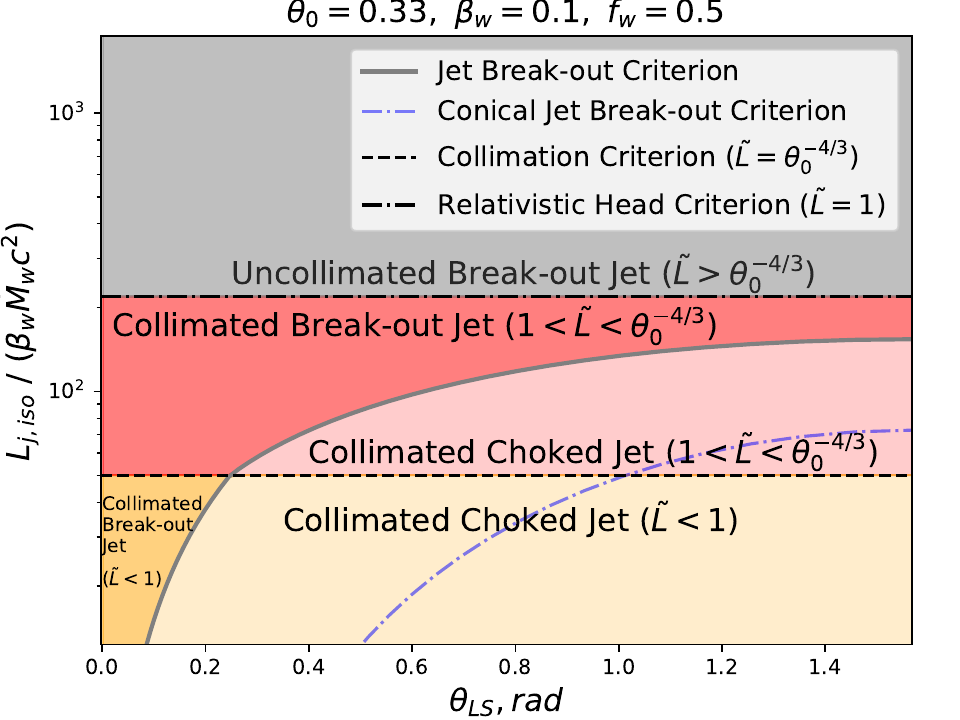}
    \caption{Break-out condition in terms of the dimensionless term $L_{\rm j,iso} / \beta_{\rm w} \dot{M}_{\rm w} c^2$ and misaligned angle $\theta_{\rm LS}$ using the same parameters as those in Fig. \ref{Lj}. The solid gray line represents the break-out criterion when the  the collimation caused by the cocoon pressure is taken in into account, while the blue dot-dashed line shows the break-out condition without taking into account the collimation effect (i.e., assuming that jets are conical). The meanings of the regions with different colors are the same as that in Fig.\ref{Lj}.}
    \label{Ljiso}
\end{figure}

For an easy comparison with the results in \cite{Lu2023arXiv231015336L}, we  express the break-out condition using the isotropic luminosity $L_{\rm j,iso}$  by using Eq.\eqref{Thetaj} and Eq.\eqref{boc2}:
\begin{align}
\begin{split}
\frac{L_{\rm j,iso}}{\beta_{\rm w} \dot{M}_{\rm w} c^2}>\left \{ 
\begin{array}{ll}
    (2^{-1/2}\beta_{\rm w}^{1/2}\pi^2{\rm sin}^2\theta_{\rm LS}\theta_0^{-4})^{2/3}/\beta_{\rm w},                    & \tilde{L} < 1\\
    (2^{-3/2}\beta_{\rm w}^{-1/2}\pi^2{\rm sin}^2\theta_{\rm LS}\theta_0^{-4})^{2/5}/\beta_{\rm w},                    & 1< \tilde{L} < \theta_0^{-4/3}\\
    \pi^2 \rm sin^2 \theta_{\rm LS}\theta_0^{-2},       &\tilde{L}>\theta_0^{-4/3}
\end{array}
\right.
\end{split}
\label{ljiso}
\end{align}
The corresponding break-out condition and various regimes of the jet  are shown in Fig.\ref{Ljiso}. The gray solid line represents the break-out criterion when the jet collimation effect is taken into account, while the blue dot-dashed line represents the case assuming a conical jet structure. A much larger $L_{\rm j,iso}$ is allowed for choked precessing jets when the jet collimation effect is properly taken into account.  




\section{High-energy neutrinos from choked precessing jets}
The shocks in choked jets will accelerate cosmic rays to ultra-high energies \citep{Farrar2009ApJ...693..329F,Wang2011PhRvD..84h1301W,Wang2016PhRvD..93h3005W} and further produce high-energy neutrinos \citep{Wang2016PhRvD..93h3005W,Mukhopadhyay2023arXiv230902275M,Lu2023arXiv231015336L}.
Internal shocks may occur due to the internal collisions within the jets, resulted from the inhomogeneity in the jet velocity.
Temporal variability with $\delta t\approx100s$ has been seen in the X-ray emission of the jetted TDE Swift J1644+57, which is thought to be generated by internal shocks \citep{Burrows2011Natur.476..421B}. We take a fiducial physical value $\Gamma \gtrsim 1/\theta_0$
for the jet Lorentz
factor, and hence the collision occurs at $R_{\rm int}\approx\Gamma^{2}c\delta t=3\times10^{14}{\rm cm}\Gamma_1^2$. 
Internal shocks that propagate into the low-density jets are collisionless,
although they locate inside the optically thick envelope \citep{Wang2016PhRvD..93h3005W}. 


One important process for neutrino production in choked jets is the interaction between the cosmic rays and the thermal photons generated in the wind envelope (i.e., the $p\gamma$ process). The cooling timescale of $p\gamma$ process is
\begin{equation}
    t^{-1}_{p\gamma} = \frac{c}{2 \gamma_p^2} \int_{\epsilon_{\text{th}}}^{\infty} \sigma_{p\gamma}(\bar{\epsilon}) \kappa_{p\gamma}(\bar{\epsilon}) \bar{\epsilon} \, d\bar{\epsilon} \int_{\bar{\epsilon}/2\gamma_p}^{\infty} \epsilon^{-2} \frac{dn}{d\epsilon}d\epsilon ,
    \label{pgamma timescale}
\end{equation}
   where $\gamma_p$ is Lorentz factors of protons, ${dn}/{d\epsilon}$ is the number density of seed photons, $\epsilon_{\text{th}} \simeq 145 {\rm MeV}$ is the threshold energy of $p\gamma$ interaction, $ \sigma_{p\gamma}$ is cross-section of $p\gamma$ interaction and $\kappa_{p\gamma}$ is inelasticity. To calculate the neutrino flux, we need to know the  the density of thermal photons in the jet. Since the jet size is much smaller than that of the wind envelope, the density of thermal photons around the jet should be roughly equal to the density of thermal photons in the envelope at the same radius. We use the same treatment of thermal photons as that in \cite{Zheng2023ApJ...954...17Z}. 
The photospheric temperature can be obtained from observations of TDEs.    When the diffusion timescale is shorter than the dynamic timescale, photons escape from the optically thick envelope through diffusion. Then from the energy
transferring equation, we can derive the temperature inside the envelope, which follows $T\propto r^{-(n+1)/4}$ for $r\ll R_{\rm ph}$ \citep{Roth2016ApJ...827....3R}, where $R_{\rm ph}$ is the photosphere radius. For the quasi-isotropic wind , we have $n=2$ and $T\propto r^{-3/4}$.    On the other hand, when the diffusion timescale exceeds the dynamic timescale, photons become trapped within the fluid, leading to that the radiation temperature is determined only by the adiabatic cooling. In this case, the temperature of wind evolves as $T \propto r^{-2/3}$ for quasi-isotropic wind with $n=2$  \citep{Piro2020ApJ...894....2P,Matsumoto2021MNRAS.502.3385M}. The typical energy of protons interacting with thermal photons with temperature $T=10^5-10^6 {\rm K}$ is $\epsilon_{\rm p}=0.1{\rm GeV}^2/3kT=0.6-6{\rm PeV}$ and the neutrino energy is correspondingly $\epsilon_{\nu}=0.05\epsilon_{\rm p}=30-300{\rm TeV}$,  consistent with the measured energy of three neutrinos possibly associated with TDEs.

   Considering that  the duty cycle of the neutrino radiation cone for a  precessing jet is $\chi_{\rm duty}=\theta_b/\pi {\rm sin}\theta_{\rm LS}$, where $\theta_b ={\rm max} \{\theta_{\rm j},1/\Gamma\}$ is the half opening angle of neutrino radiation cone. The fluence of neutrinos   from this precessing jet is
\begin{equation}
    \epsilon_{\nu}\mathcal{F}_{\nu}\approx\int^{t_{\rm d}/(1+z)}_{0}\frac{L_{\rm p}f_{\rm p\gamma} e^{-\frac{\epsilon_{\rm p}}{\epsilon_{\rm p,max}}}}{32\pi D^2_{\rm L}\ln(\epsilon_{\rm p,max}/\epsilon_{\rm p,min})} \chi_{\rm duty} dt ,
    \label{flu}
\end{equation}
where $t_d$ is the neutrino trigger time,  $L_{\rm p}=\varepsilon_{\rm p}L_{\rm iso}$ is the isotropic proton luminosity which is defined as $L_{\rm iso} = 2L_{\rm j}/\theta_b^2$, $\varepsilon_{\rm p}$ is the fraction of the jet energy converted into relativistic protons  and $\Gamma$ is the bulk Lorentz factor of the jet. $f_{\rm p\gamma}$ is the  fraction of energy loss by the $p\gamma$ process, 
   which can be expressed as $t_{\rm p\gamma}^{-1}/(t_{\rm p\gamma}^{-1}+t_{\rm dyn}^{-1})$, where $t_{\rm dyn} \approx R_{\rm int}/c$ is dynamical timescale of the jet. In our calculation  $f_{\rm p\gamma} \simeq 1$ is obtained. Assuming that the jet power scales with the accretion rate, the jet luminosity evolves with time as 
$L_{\rm j}\propto (t/\tau)^{-\frac{5}{3}}$ after the peak  \citep{Krolik2012ApJ...749...92K,Piran2015MNRAS.453..157P},  where  $\tau$ is the  characteristic fallback timescale, which is roughly the orbital period of the most bound debris \citep{Gezari2021ARA&A..59...21G}, i.e.,
\begin{equation}
    \tau=41M^{1/2}_{\rm BH,6}\left(\frac{M_{\star}}{M_\odot}\right)^{-0.1}{\rm day},
    \label{tau}
\end{equation}
with $M_{\rm BH}$ being the mass of SMBH and $M_{\star}$ being the mass of the disrupted star.  
   The maximum energy of protons is obtained by equating the acceleration time $t_{\rm acc} = \gamma_{\rm p}m_{\rm p}c/\eta_{\rm acc}eB$ to the cooling time $t_{\rm p\gamma}$, where $\eta_{\rm acc} = 0.1$ is acceleration efficiency, and magnetic field is given by $B = \sqrt{4\varepsilon_{\rm B}L_{\rm j}\Gamma^2/R_{\rm int}^2c}$. We take the minimum energy of protons as $\epsilon_{\rm p,min}=\Gamma m_{\rm p}c^2$ and assume  equipartition factors $\varepsilon_{\rm B}=0.1$ and $\varepsilon_{\rm p}=0.2$ in the calculation.
Then we can obtain the cumulative muon neutrino number between 10 TeV and 1PeV by using
\begin{equation}
    N_{\nu}=\int^{1 {\rm PeV}}_{10 {\rm TeV}}A_{\rm eff}(\epsilon_{\nu})\mathcal{F}_{\nu}d\epsilon_{\nu}.
    \label{Numb}
\end{equation}
where $A_{\rm eff}$  is the   effective area of IceCube\citep{Blaufuss2019ICRC...36.1021B}.

\begin{figure}
     \centering
    \includegraphics[width=0.5\textwidth]{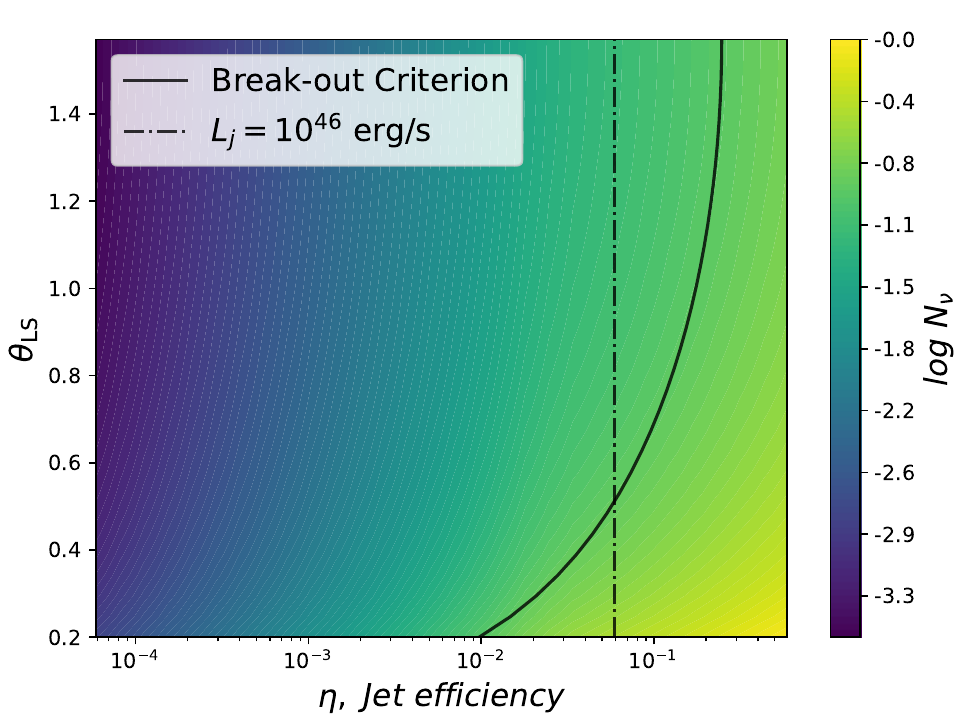}
    \caption{Color map of neutrinos numbers as the function of misaligned angle versus jet power in a general case. The solid black line represents the breakout condition, and the dot-dashed black line corresponds to a jet similar to Swift J1644 with a jet luminosity of $10^{46} \rm erg ~s^{-1}$. The left side of the black solid line corresponds to the region that jets will not break through the envelope and vice versa.}
    \label{General} 
\end{figure}

\begin{figure*}
\centering
\subfigure[]{
\includegraphics[width=0.32\textwidth]{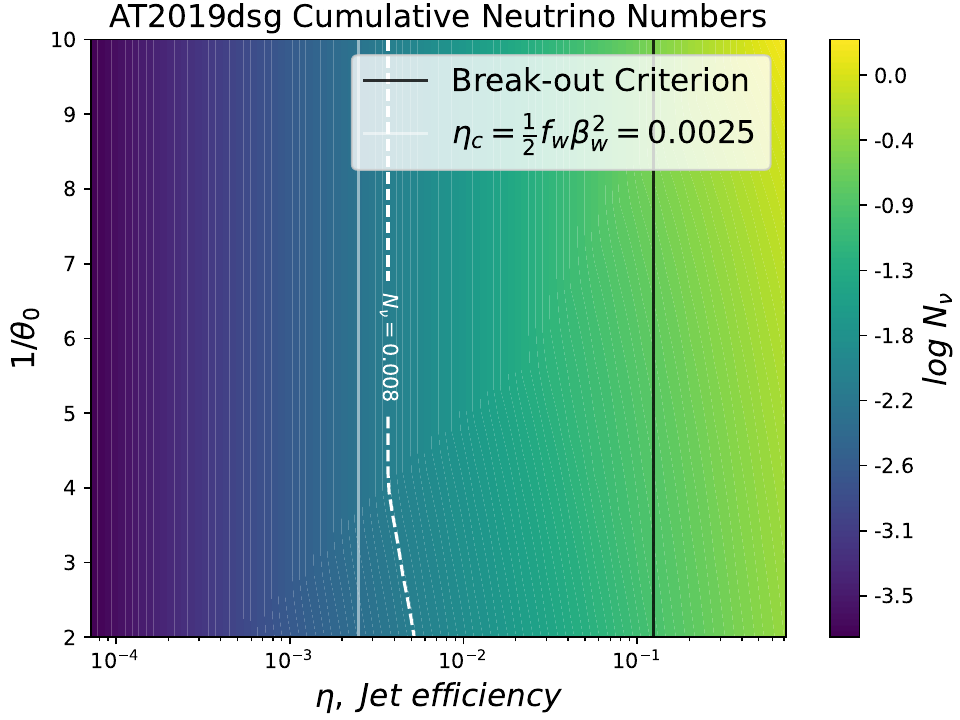}}
\subfigure[]{
\includegraphics[width=0.32\textwidth]{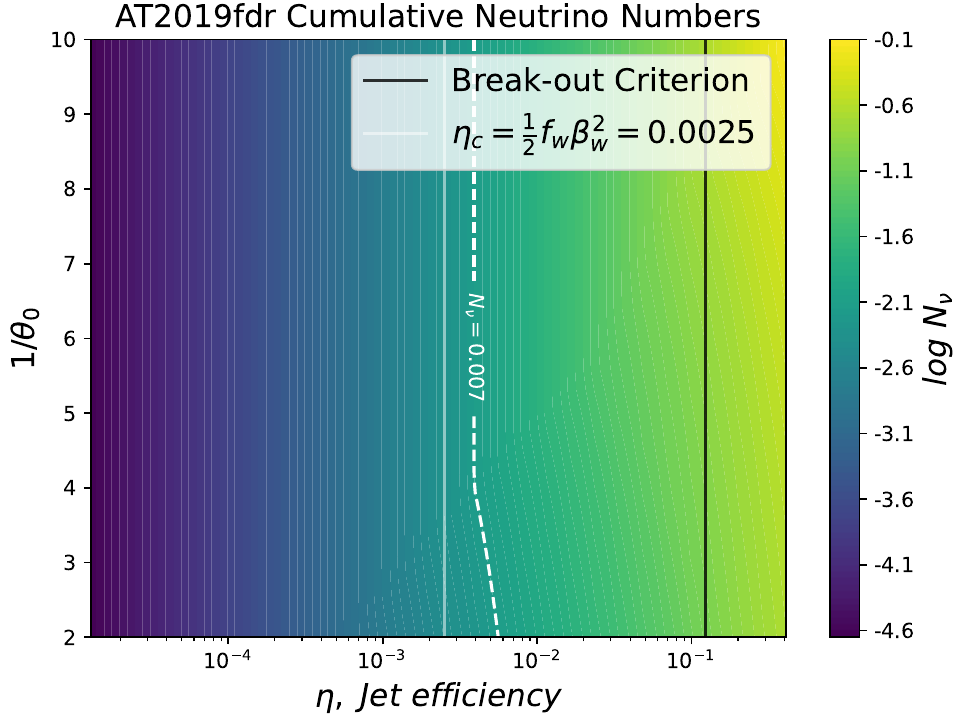}}
\subfigure[]{
\includegraphics[width=0.32\textwidth]{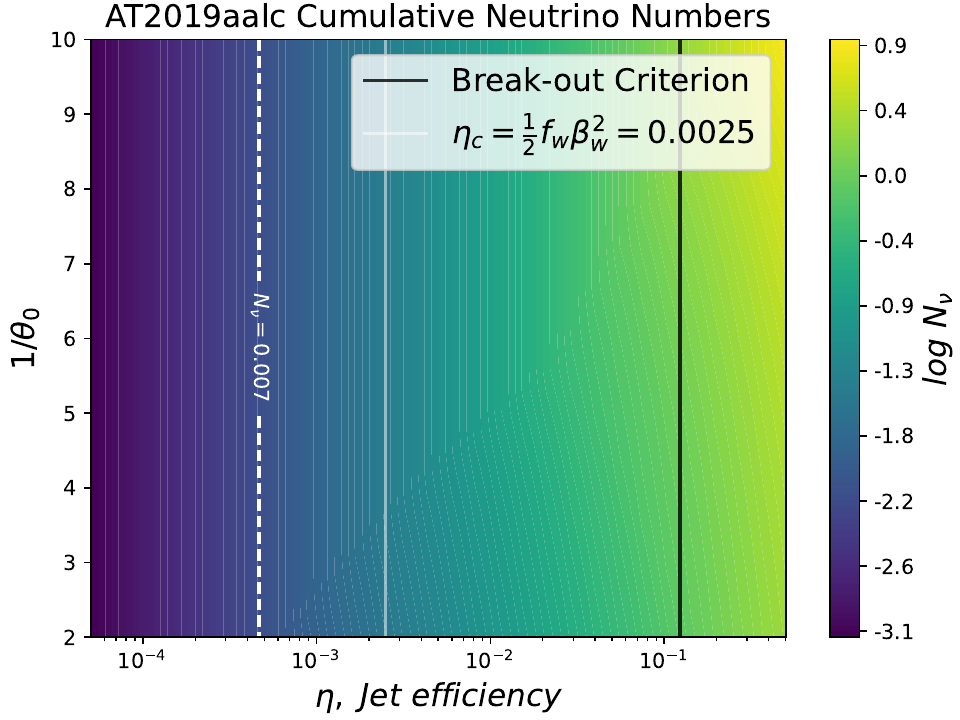}}
\caption{Color map of the expected neutrino numbers  for three TDEs (AT2019dsg, AT2019fdr, AT2019aalc) as a function of the jet efficiency $\eta$ and  initial  opening angle $\theta_0$ with a fixed misaligned angle $\theta_{\rm LS} = \pi/4$. The white dashed line represent the minimal required neutrino numbers inferred from observations.  The  black vertical line represents the maximum jet efficiency limited by the break-out criterion, while the white solid line represents the critical jet efficiency $\eta_c = 0.0025$.}
\label{colormap Gamma-Lj}
\end{figure*}

\begin{figure*}
\centering
\subfigure[]{
\includegraphics[width=0.32\textwidth]{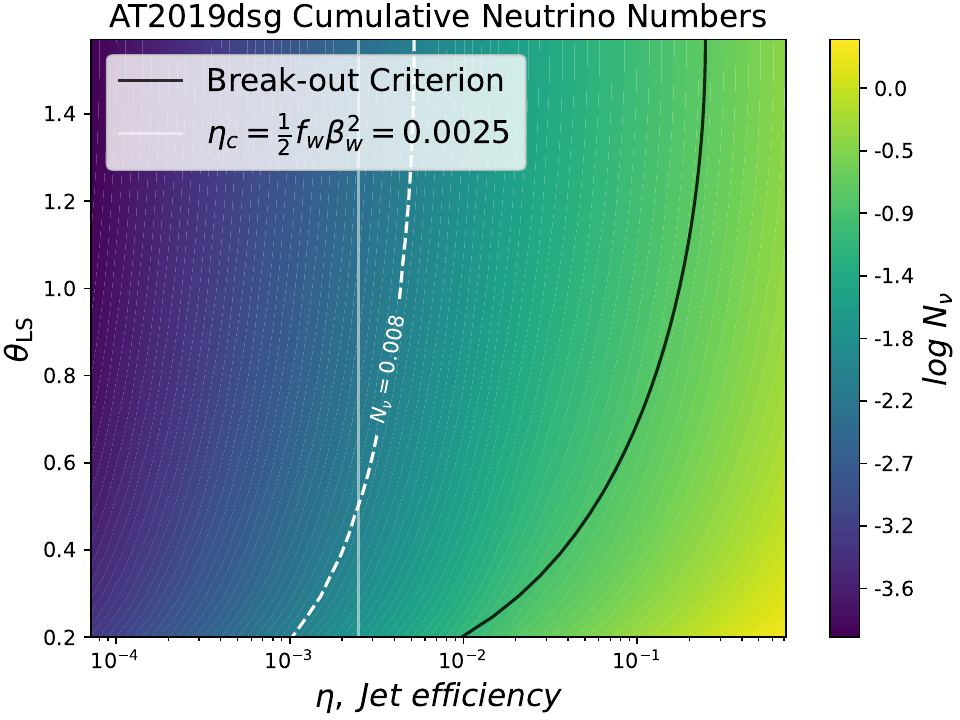}}
\subfigure[]{
\includegraphics[width=0.32\textwidth]{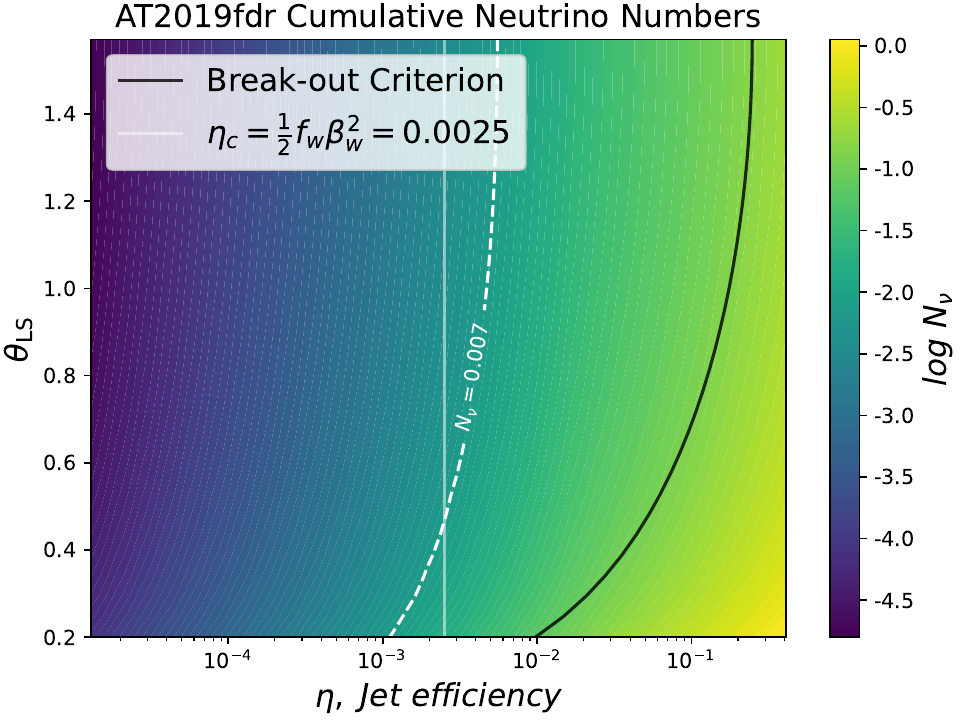}}
\subfigure[]{
\includegraphics[width=0.32\textwidth]{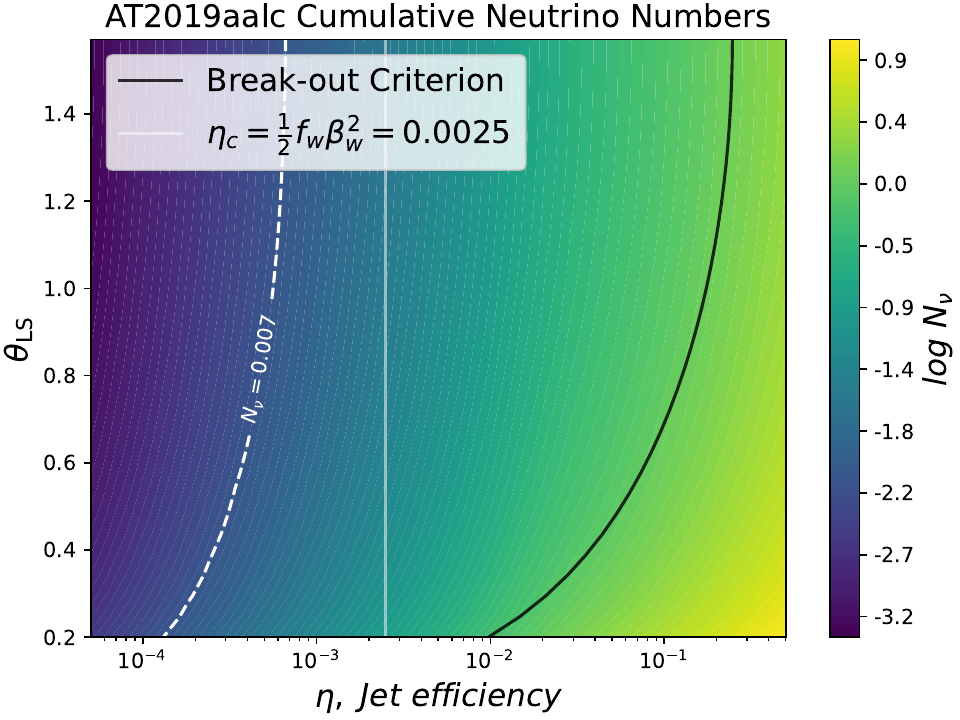}}
\caption{Color map of the expected neutrino numbers   for three TDEs (AT2019dsg, AT2019fdr, AT2019aalc) as a function of the jet efficiency $\eta$ and misaligned angle $\theta_{\rm LS}$ with initial jet half-opening angle of $\theta_{0} = 0.33$ and $\Gamma = 10$. The meanings of each lines are the same as that in Fig.\ref{colormap Gamma-Lj}.}
\label{colormap Psi-Lj}
\end{figure*}

We calculate the expected neutrino number from precesing jets as the function of the misaligned angle and jet power, as shown in Fig.\ref{General}. We take fiducial values of $\Gamma=10$, $\theta_0 = 0.33$ and $\tau = 41 {\rm day}$ (corresponding to a SMBH with $10^6M_\odot$) for the jets.  We take a reference value ($D={\rm  220Mpc}$) for the TDE distance, equal to the distance of AT2019dsg. As expected, the neutrino flux increases as the jet power increases. The neutrino flux also increases  as the misaligned angle decreases,   because the  duty cycle  that the
jet sweeps across increases as the misaligned angle decreases.  The black solid line represents the break-out condition given by Eq.\eqref{bocj}. The maximum neutrino number that  precessing choked TDEs can produce is $N_{\nu}\sim 0.1$. The black dot-dashed line shows the case of TDEs like Swift J1644+57, which has a high luminosity of $10^{46} \rm erg ~s^{-1}$. For this case, the maximum neutrino number produced by the choked jet is about $N_{\nu}\sim 0.1$.

\subsection{ Could the neutrinos from AT2019dsg, AT2019fdr and AT2019aalc be produced by choked  precessing jets? }
We explore the parameter space of the precessing jets that can produce detectable neutrino flux in three TDE candidates (AT2019dsg, AT2019fdr and AT2019aalc)  while being choked in the wind envelope. 
To calculate the maximum neutrino flux in the
three TDEs, we need  to know the maximum  luminosity allowed for choked jets, which can be obtained from Eq.\eqref{bocj}. Meanwhile,  the  maximum  luminosity could also be limited by the maximally  allowed deposited energy by choked jets, since this deposited energy will eventually transfer to the wind envelope via the cocoon.  A detailed numerical study is needed to understand how the deposited energy by choked jets affects the wind envelope and the radiation. For simplicity, we assume that the cocoon energy does not exceed the kinetic energy of the disk wind, otherwise the kinetic energy of the disk wind will increase significantly.
It will be shown below that the critical jet luminosity limited by the maximally allowed  deposited energy by choked jets is much smaller than the jet luminosity  limited by the jet break-out condition for the three TDEs,  so the real constraints on the neutrino flux come from the maximally allowed  deposited energy.



The mass of the disrupted stars in three TDEs is not well-known. Similar to \cite{Zheng2023ApJ...954...17Z},  we assume the mass of the  disrupted stars   scales with the bolometric energy, i.e., $E_{\rm bol} \propto M_{\star}$ \citep{Metzger2016MNRAS.461..948M}. Then the  masses of the  disrupted stars are estimated to be $0.8 M_{\odot}$, $20 M_{\odot}$ and  $4 M_{\odot}$ for AT2019dsg,  AT2019fdr and AT2019aalc, respectively. Another important parameter is the mass-loss rate $\dot{M}_{\rm w}=f_{\rm w}\dot{M}$ in the disk wind, where $\dot{M}=M_{\star}/(3\tau)$ is the mass fallback rate. The characteristic fallback timescale $\tau$ for three TDE AT2019dsg, AT2019fdr, AT2019aalc are  40 days \citep{Velzen2021ApJ...908....4V}, 180 days and 140 days  calculated by Eq.\eqref{tau} respectively. 

We assume that the deposited cocoon energy should not exceed the kinetic energy of the disk wind too much, otherwise the wind kinetic energy will be significantly increased.
The power of wind is represented by $L_{\rm w} = \frac{1}{2}\dot{M}_{\rm w} v_{\rm w}^2$. In the following calculation, we set $\beta_{\rm w} = 0.1$ and $f_{\rm w}=0.5$. 
For AT2019dsg, the maximum jet luminosity limited by a choked jet can reach $L_{\rm j}=1.7 \times 10^{46}~{\rm 
 erg~s^{-1}}$. 
However, the wind kinetic power of AT2019dsg is estimated to be only $3.5\times 10^{44}~{\rm erg~s^{-1}}$, which actually limits the allowed maximum jet luminosity. 
Similarly, for AT2019fdr, the  maximum luminosity of a choked jet reaches $L_{\rm j}=9.5 \times 10^{46}~{\rm erg~s^{-1}}$, while the wind kinetic power is  $1.9\times 10^{45}~{\rm erg~s^{-1}}$.
For AT2019aalc, the maximum jet luminosity for a  choked jet is $L_{\rm j}=2.5 \times 10^{46}~{\rm erg~s^{-1}}$, while the wind kinetic luminosity is $4.9\times 10^{45}~{\rm erg~s^{-1}}$.

The deposited cocoon energy can be expressed as 
\begin{equation}
E_c = \int L_{\rm j} (1-\beta_{\rm h}) dt .
\label{Ec}
\end{equation}
If we consider the wind power for three TDEs as the  jet luminosity, the cocoon energy, using Eq.\eqref{betahead} and Eq.\eqref{Ec}, is estimated to be $1.4 \times 10^{51}~{\rm erg}$, $3.5 \times 10^{52}~{\rm erg}$ and $7\times 10^{51}~{\rm erg}$ for AT2019dsg, AT2019fdr and AT2019aalc, respectively.

The expected neutrino numbers of three TDEs are shown by the color intensity in Fig.\ref{colormap Gamma-Lj}.  The dashed white lines show the parameter values corresponding to the minimum neutrino number of each TDE required by observations at  $90\%$ confidence\citep{Stein2021NatAs...5..510S}.  The solid black lines show the critical jet efficiency limited by the break-out condition and the solid white lines show the  jet efficiency limited by the power of the disk wind.    The break-out criterion is normalized based on jet efficiency as $\eta \ge \frac{1}{2}f_{\rm w}\beta_{\rm w}\pi^2{\rm sin}^2\theta_{\rm LS}$. Meanwhile the critical jet efficiency $\eta_c = \frac{1}{2}f_{\rm w}\beta_{\rm w}^2 = 0.0025$.
Fig.\ref{colormap Gamma-Lj} shows that precessing choked jet model can explain the neutrinos from AT2019aalc, while it can only marginally explain the neutrinos from  AT2019dsg and AT2019fdr for a fixed large misaligned angle $\theta_{\rm LS} = \pi/4$. 
Fig.\ref{colormap Psi-Lj} illustrates the impact of the misaligned angle $\theta_{\rm LS}$ on the expected neutrino number. A weakly misaligned angle can result in a high duty cycle, consequently increase the neutrino flux. 
For AT2019dsg and AT2019fdr, misaligned angles less than 0.4 radians can meet the requirement of the observed neutrino flux. The most optimistic estimate of the neutrino number in the weakly misaligned jet scenario can reach $N_{\nu} = 0.02$, $N_{\nu} = 0.015$ and $N_{\nu} = 0.12$ for AT2019dsg, AT2019fdr and AT2019aalc respectively. 

\subsection{Diffuse neutrino emission from Choked TDE jets }
If choked precessing jets are common in normal TDEs, neutrinos from these choked precessing jets may contribute to the all-sky diffuse neutrino background.
A rough estimate of the diffuse muon neutrino emission from these jets, using the above constraints and the updated event rate of normal TDEs \citep{Yao2023ApJ...955L...6Y}, can be obtained by
\begin{equation}
\begin{aligned}
    \varepsilon^2_{\nu}\Phi_{\nu} &\approx\frac{c}{{4\pi {H_0}}}f_{\rm z} E_\nu \rho_0 
      \\
    &\approx3\times10^{-9} f_{\rm z}f_{\rm p\gamma}E_{\rm p,51}\left(\frac{\rho_0}{310{\rm Gpc^{-3}\,yr^{-1}}}\right)\\
    &\quad{\rm GeV\,cm^{-2}\,s^{-1}\,sr^{-1}},
\end{aligned}
\label{diff}
\end{equation}
where $E_{\nu}=f_{\rm p\gamma}E_{\rm p}/8\ln(\epsilon_{\rm p,max}/\epsilon_{\rm p,min})$ is the muon neutrino energy released by a single TDE event,
$E_{\rm p}$ is total CR proton energy injected by one TDE event, $\rho_0$ is the local event rate of TDEs, $H_0$ is the Hubble constant, and $f_{\rm z}$ is the correction factor for the contribution from high redshift TDEs, which is uncertain from current observations. 
For the TDE event rates, a latest demography study suggested that the local TDE event rate is $\rho_0\simeq 310{\rm Gpc^{-3}\,yr^{-1}}$ \citep{Yao2023ApJ...955L...6Y}. A typical solar mass TDE can eject half of the mass into the disk wind at a velocity $v_{\rm w}\simeq 0.1{\rm c}$\citep{Metzger2016MNRAS.461..948M}, leading to a kinetic energy of $E_{\rm k}\approx2.3\times10^{51}(f_{\rm w}/0.5)(v_{\rm w}/0.1{\rm c})^2$ erg in the disk wind. Noting that this energy does not violate the constraints from radio observations of some TDEs, as the ejecta is still in the free-expansion phase\citep{2021MNRAS.507.4196M}. Radio observations show that the kinetic energy of the jets in Swift J1644+57 and AT2022cmc is about $E_{\rm k}\sim 10^{51}-10^{52}{\rm erg}$ \citep{Cendes2023arXiv230813595C}.  This motivated us to assume that the kinetic energy of the choked jet is comparable to  the kinetic energy of the disk wind,  so the CR proton energy in one TDE event is $E_{\rm p}\sim  4.5\times 10^{50}(v_{\rm w}/0.1{\rm c})^2$ erg if taking $\epsilon_{\rm p}=0.2$. Assuming $f_{\rm p\gamma}\simeq 1$ and $f_{\rm z}\simeq 1$,  the single flavor neutrino  flux contributed by choked TDE jets is approximately $\sim 10^{-9}{\rm GeV\,cm^{-2}\,s^{-1}\,sr^{-1}}$, which is about $\sim 10\%$ of the diffuse neutrino flux at 30-300 TeV \citep{Abbasi2022ApJ...928...50A}.



\section{Summary and Discussions}
We revisited the jet break-out condition for misaligned precessing jets in TDEs by considering that the jet could be collimated by the cocoon pressure while the jet is propagating in the  quasi-isotropic disk wind. We find that the collimation effect is important for choked precessing jets.  As a result, the jet head opening angle decreases as the jet propagates in the envelope, and the isotropic jet luminosity $L_{\rm j,iso}$ increases correspondingly.  Nevertheless, the break-out condition does not change if we consider the beam-corrected jet luminosity, which is simply  $L_{\rm j}\ge \frac{1}{2}\dot{M} f_{\rm w} c^2 \beta_{\rm w}\pi^2 {\rm sin}^2\theta_{\rm LS}$. Thus, for known values of $f_{\rm w}$, $\beta_{\rm w}$ and $\theta_{\rm LS}$, the break-out condition depends only on the jet efficiency $\eta=L_{\rm j}/\dot{M}  c^2$. For typical parameter values of $f_{\rm w}=0.5$, $\beta_{\rm w}=0.1$ and $\theta_{\rm LS}=\pi/4$, successful jets imply $\eta> 10^{-1}$. 

   An important result of \citet{Lu2023arXiv231015336L} is that a successful jet breakout requires a double alignment condition, which naturally explains why the event rate of jetted TDEs is much smaller than optical/X-ray TDEs.  They find that the fraction of jetted TDEs that meet the double alignment condition is $\sim 10^{-3} ({\theta_j}/0.1)^4$ assuming a half jet opening angle of $\theta_j=0.1$. The $\theta_j=0.1$ case of a conical jet in \citet{Lu2023arXiv231015336L} corresponds to   $\theta_0\sim 0.33$ in our case, which has a half jet opening angle  $0.1$ after collimation. 

We further study the neutrino emission from  choked precessing  jets. Protons accelerated by shocks inside the choked precessing jet interact with the dense thermal photons in the wind envelope and produce high-energy neutrinos. In Fig.\ref{colormap Gamma-Lj} and Fig.\ref{colormap Psi-Lj} we show  the cumulative neutrino numbers for three TDEs in the parameter space of  $\theta_0-L_{\rm j}$ and $\theta_{\rm LS}-L_{\rm j}$. Although precession allows a much larger jet luminosity before break-out  (compared with the aligned jet case discussed in \cite{Zheng2023ApJ...954...17Z}), the real physical jet power can not be so large, otherwise the deposited energy by choked jets would exceed the wind power.  For the same jet luminosity, the observed neutrino flux from a choked precessing  jet is  reduced by a factor represented by the duty cycle due to precession, which is $\theta_0/(\pi  \sin\theta_{\rm LS})$. We propose that the neutrinos from AT2019dsg, AT2019fdr are more likely to be produced by weakly misaligned choked jets, and neutrinos from AT2019aalc are probably emitted by misaligned choked precessing jets.
We also calculate the all-sky diffuse neutrino flux from all choked TDE jets and find that it constitutes a fraction of $\sim10\%$ of the total neutrino background flux.

   If the jet luminosity is limited by the wind power, a majority of TDE jets may have lower efficiency than observed jetted TDEs.  We assume the  critical value $\eta_c$ as a typical choked jet efficiency, then the dimensionless black hole spin parameter is $a \sim 0.2-0.3$ by utilizing $\eta = 4\times 10^{-3} \Phi_{\rm B}^2\Omega_{\rm H}^2(1+1.38\Omega_{\rm H}^2-9.2\Omega_{H}^4)$, where $\Omega_{\rm H} = a/(1+(1-a^2)^{1/2})$ is the dimensionless angular frequency of the horizon and $\Phi_{\rm B} \simeq -20.2a^3-14.9a^2+34a+52.6$ is dimensionless magnetic flux threading the black hole's horizon\citep{Teboul2023ApJ...957L...9T,Lu2023arXiv231015336L}. 

In this work, we assumed that the energy of the choked jet is limited by the kinetic energy of the disk wind. However, it is unknown how the deposited energy by choked jets affects the wind envelope and the radiation. A detailed numerical study is needed to verify this assumption. If the maximally allowed energy for the choked jets is larger, the neutrinos flux will become larger correspondingly. 


\begin{acknowledgments}
We thank Wenbin Lu and Weihua Lei for valuable discussions. The work is supported by the NSFC under grants Nos. 12333006, 12121003 and 12393852. 
\end{acknowledgments}

\bibliography{reference}{}
\bibliographystyle{aasjournal}

\end{document}